\documentclass[doublecol]{epl2}

\usepackage{amsfonts,amssymb,amsthm,bbm,amsmath}

\usepackage{color}

\newcommand{\C}{{\mathbb C}}

\newcommand{\R}{{\mathbb R}}

\newcommand{\cH}{{\mathcal H}}

\newcommand{\cO}{{\mathcal O}}

\newcommand{\cC}{{\mathcal C}}

\newcommand{\be}{\begin{equation}}
\newcommand{\ee}{\end{equation}}
\newcommand{\beq}{\begin{eqnarray}}
\newcommand{\eeq}{\end{eqnarray}}
\newcommand{\bes}{\begin{eqnarray}}
\newcommand{\ees}{\end{eqnarray}}

\newcommand{\mat} [2] {\left ( \begin{array}{#1}#2\end{array} \right ) }
\newcommand{\bino} [2] {\mat{c}{#2 \\ #1}}

\renewcommand{\sl}{{\mathfrak{sl}}}

\newcommand{\la}{\langle}
\newcommand{\ra}{\rangle}

\newcommand{\f}{\frac}

\def\nn{\nonumber}
\def\pp{\partial}

\def\rd{\mathrm{d}}

\def\ka{\kappa}

\def\om{\omega}

\def\hx{\hat{x}}
\def\hp{\hat{p}}
\def\hD{\widehat{D}}
\def\hH{\widehat{H}}

\def\op{\cO}

\def\bpsi{\bar{\psi}}

\title{Evolution of the wave-function's shape in a time-dependent harmonic potential}

\author{E.R. Livine\inst{1,2}}

\institute{                    
  \inst{1} Universit\'e de Lyon, ENS de Lyon, CNRS, Laboratoire de Physique LPENSL, 69007 Lyon, France\\
  \inst{2} Perimeter Institute for Theoretical Physics,  Waterloo, Ontario N2L 2Y5, Canada
}

\abstract{
An effective operational approach to quantum mechanics is to focus on the evolution of wave-packets, for which the wave-function can be seen in the semi-classical regime as representing a classical motion dressed with extra degrees of freedom describing the shape of the wave-packet and its fluctuations. These quantum dressing are independent degrees of freedom, mathematically encoded in the higher moments of the wave-function.
We review how to extract the effective dynamics for Gaussian wave-packets evolving according to the Schr\"odinger equation with time-dependent potential in a 1+1-dimensional spacetime, and derive the equations of motion for the quadratic uncertainty. We then show how to integrate the evolution of all the higher moments for a general wave-function in a time-dependent harmonic potential. 
}

\begin{document}
\maketitle



The wave-function is the mathematical description of systems in quantum mechanics. Putting aside  measurement processes and eigenstate projections, its dynamics is formulated as a standard classical space-time field following the non-relativistic Schr\"odinger equation. It is customary to focus on Fourier modes, especially because they are exact solutions of the free Shr\"odinger equation and furthermore asymptotic solutions to the Shr\"odinger equation with an asymptotically-vanishing potential. So it is the natural basis to define and study the S-matrix. However, it is also physically revealing to look instead at the evolution of the moments of the wave-functions, defined as the expectation values of the powers of the position and momentum operators. These characterize the localization, spread, shape and fluctuation of the wave-function. 

An operational approach to quantum mechanics is then to derive effective equations of motion for those moments (see \cite{Bojowald:2005cw,Bayta2019,Bojowald:2022lbe} and references therein). As a first approximation, as a first dive into quantum phenomenology at leading order, one looks at the evolution of Gaussian wave-packets (see e.g.\cite{Heller,ABCV, Pattanayak_1994,Prezhdo, Blum:1995wi}) and derive equations of motion for the position, momentum and quadratic quantum uncertainties. This illustrates the semi-classical perspective of quantum mechanics as dressed classical mechanics with extra modes encoding the fluctuations of shape of the wave-function. Then one wants to go further, beyond Gaussian wave-packets, and study the whole tower of higher modes.
This effective approach is naturally relevant to the study of the quantum-to-classical transition. 

Here we write the evolution equations for all the moments of the wave-function, according to Schr\"odinger equation, without truncating them to the second order in the position and momentum operators. We focus on a harmonic potential, though we allow it to be explicitly time dependent. In this case with no anharmonic potential terms, the higher modes are dynamical, but decouple from lower modes and do not interfere with their dynamics. More precisely, a  harmonic potential implies that moments $\la\hx^a\hp^b\ra$ decouple according to their degree $n=a+b$ in $\hx$ and $\hp$, leading to a structure for those polynomial observables with layers corresponding to fixed order $n$. We identify the space of solutions of equations of motion for each layer, check that it has the expected dimension of $(2n+1)$, and show that a basis of solutions can be directly constructed from polynomials in the two solutions for the classical trajectory. Thus although all the moments evolve independently, they all beat with the same rhythm. It shows explicitly a synchronization of the wave-function fluctuations, across all multipole moments. This is rather surprising that the time dependence of the potential does not interfere with this behavior.
This exact solution for the time dependent quantum harmonic oscillator, together with the classical analysis in \cite{Livine:2022vaj}, completes other existing studies, e.g. \cite{Kanasugi,PhysRevResearch.2.043162,fiore2022timedependent}.
Indeed, an exact solution for the evolution operator has been known for some time, from which one can extract the evolution of the quadratic moments, i.e. the squeezing behavior of the wave-function, e.g. \cite{PhysRevA.43.404}. It can however be a tedious task to derive the evolution of the whole tower of higher moments from a formal expression of the unitary evolution operator. Here, we present an elementary derivation of the dynamics of those higher moments, by deriving the equations of motion directly satisfied by those moments and providing the explicit solution to these equations. This leads to a clear picture of the evolution of the shape of wave-packets in a time dependent harmonic potential. 

This simple exercise is meant as a springboard towards a more general analysis of the evolution of quantum uncertainty for non-linear extensions of Schr\"odinger equations, where the self-interaction of the wave-function will lead to a non-trivial coupled dynamics for quantum superpositions of wave-packets. This could for instance be applied to the phenomenology of quantum physics of gravitating systems
\footnote{
Starting from 1+1-d Schr\"odinger action for a non-relativistic massive scalar system,
\be
S[\psi]=\int \rd t\rd x\,
\big{[}
i\hbar \bpsi\pp_{t}\psi
-\f{\hbar^{2}}{2m}\pp_{x}\bpsi\pp_{x}\psi
\big{]}
\,,\nn
\ee
we recall the physical dimensions of the wave-function $[\psi]^{2}=L^{-1}$ and of Planck constant $[\hbar]=ML^{2}T^{-1}$. If one were to consider gravitational corrections as possible interaction terms  at leading order with the Newton constant as coupling constant, considering its physical dimension $[G]=M^{-1}L^{3}T^{-2}$ and reasonably restricting ourselves to real translationally-invariant polynomial terms, it appears that only two gravitational couplings are possible,
\be
\Delta S_{G}[\psi]=\int \rd t\rd x\,
\bigg{[}
\alpha  Gm^{2} i(\bpsi\pp_{x}\psi-\psi\pp_{x}\bpsi)
+\beta Gm^{2}|\psi|^{4}
\bigg{]}
\,.\nn
\ee
The first is the translational flux, while the second leads to the simplest non-linear Schr\"odinger equation.
}
, which has become an active topic of research with possible lab experiments, e.g.\cite{Belenchia:2019gcc,Westphal:2020okx,Christodoulou:2022mkf,Biswas:2022qto}.

\section{Wave-packet Dynamics}

Let us start with  a single 1d particle of mass $m$ evolving a time-dependent potential $V(t,x)$, whose classical mechanics is driven by the action,
\be
s[x(t)]=\int\rd t\,\left[
\f m2 \dot{x}^{2}-V(t,x)
\right]\,,
\ee
whose Euler-Lagrange equations determines the classical trajectories $x(t)$, as curves drawn in the 2d space-time:
\be 
m\ddot{x}+\pp_{x}V=0
\,.
\ee
%
Moving up to the  quantum level,  the quantum mechanics of this system us defined in terms of a wave-function $\psi(t,x)$, here chosen in the $x$-polarisation. Putting aside the questions related to measurements and the collapse of the wave-function, quantum mechanics reads as a 1+1d non-relativistic field theory driven by the following action:
\be
S[\psi]
=
\int \rd t \rd x\,
\left[
i\hbar \bpsi\pp_{t}\psi
-\f{\hbar^{2}}{2m}\pp_{x}\bpsi\pp_{x}\psi - V|\psi|^{2}
\right]\,,
\ee
whose field equation is the Schr\"odinger equation:
\be
i\hbar \pp_{t}\psi
=
\hH\psi
\,,\quad\textrm{with}\quad
\hH
=
-\f{\hbar^{2}}{2m}\pp_{x}^{2}+V(t,x)
\,.
\ee
One typically looks at momentum modes and uses the Fourier decomposition of the wave-function in order to diagonalize the Hamiltonian operator $\hH$ and determine the proper energy modes. Momentum modes are especially interesting when considering asymptotic wave modes.
Another path is to study the dynamics of wave-packets. This is the perspective of effective quantum mechanics \cite{Bojowald:2005cw,ABCV}, which considers the evolution of the wave-function as the classical motion dressed with quantum fluctuations. Then Schr\"odinger's equation gives the dynamics of the classical degrees of freedom - the position and momentum - as well as of the higher moments of the wave function, encoding the shape of the wave function and defining the extra degrees of freedom associated to the quantum particle.
This point of view is relevant to understanding the quantum-to-classical transition, since it focusses on the coupling between classical and quantum degrees of freedom and the effect of the fluctuations of the quantum uncertainty on the classical trajectory.

For instance, let us look into the dynamics of Gaussian wave-packets, defined by the wave-function ansatz:
\be
\psi_{G}^{(q,p,A)}(x,t)=Ne^{i\gamma}\,e^{\f i\hbar p(x-q)}\,e^{-A(x-q)^{2}}
\,,
\ee
where the subscript $G$ stands for Gaussian.
This state is parametrized by the pair $(q(t),p(t))\in\R^{2}$ and  the  Gaussian width $A$. We decompose the width in terms of a pair $(\alpha(t),\beta(t))\in\R^{2}$ as
\be
A=\f1{4\alpha^{2}}-\f{i}{2\hbar}\f \beta\alpha
\,.
\ee
The global phase $\gamma(t)$ evolves in time and the normalization factor $N$ is set by:
\be
1=\int |\psi|^{2}
=N^{2}\,\alpha\sqrt{2\pi}
\,.
\ee
The parameters $q,p,\alpha,\beta$ describe the kinematics of the wave-packet. Indeed they determine the expectation values of the position $\hx$ and momentum $\hp=-i\hbar\pp_{x}$, and their quadratic spread:
\beq
&&\la\hat{x}\ra=q
\,,\quad
\la\hat{x}^{2}\ra
=
q^{2}+\alpha^{2}
\,,
\\
&&\la\hat{p}\ra=p
\,,\quad
\la\hat{p}^{2}\ra=
p^{2}+\beta^{2}+\f{\hbar^{2}}{4\alpha^{2}}
\,,
\nn\\
&&\hat{D}\equiv \f12(\hat{x}\hat{p}+\hat{p}\hat{x})
\,,\quad
\la \hat{D}\ra=pq+\alpha\beta
\,.\nn
\eeq
%
%
The dynamics of the wave-packet is given by Schr\"odinger equation. Applying the evolution equation for an arbitrary observable $\cO$,
\be
i\hbar\rd_{t}\la \cO\ra
=
\la [\cO,\hH]\ra
\,,
\ee
we get Ehrenfest theorem for the first order observables:
\be
\rd_{t}\la\hat{x}\ra=\f1m\la\hat{p}\ra
\,,\quad
\rd_{t}\la\hat{p}\ra=-\la\widehat{\pp_{x}V}\ra
\,.
\ee
Keeping in mind that the expectation value $\la\widehat{\pp_{x}V}\ra$ is not the classical evaluation $\pp_{x}V|_{\la\hat{x}\ra}$ at the position expectation value, we apply those equations of motion to the Gaussian ansatz. Assuming that the potential is even and Taylor-expanding it in powers of the position,
\be
V
=V_{0}+\f{V_{2}}{2}x^{2}+\f{V_{4}}{4!}x^{4}+..
\,,\,\,
\pp_{x}V={V_{2}}x+\f{V_{4}}{6}x^{3}+..
\,,\nn
\ee
we get the effective equations of motion:
\be
\label{effdyn1}
\rd_{t}q=\f pm\,,\quad
\rd_{t}p=-\left(
V_{2}q+\f{V_{4}}{6}q^{3}+\f{V_{4}}{2}q\alpha^{2}+..
\right)\,,
\ee
where we recover that $p$ is the momentum conjugated to the position $q$, and where we identify the effective coupling in $V_{4} q\alpha^{2}$ between the classical degrees of freedom  and the quantum fluctuations due to the quartic potential terms.
Pushing the analysis to second order moment of the wave-function, we get:
\begin{align}
&\rd_{t}\la\hat{x}^{2}\ra=\f2m\la\hat{D}\ra
\,,\quad
\rd_{t}\la\hat{p}^{2}\ra=-\la\hat{p}\widehat{\pp_{x}V}+\widehat{\pp_{x}V}\hat{p}\ra
\,,
\nn\\
&\rd_{t}\la\hat{D}\ra
=
\f1m\la\hat{p}^{2}\ra-\la\widehat{x\pp_{x}V}\ra
\,.
\end{align}
Applying this to the Gaussian ansatz, we get the following equations for the quadratic uncertainty observables $(\alpha,\beta)$, here expanded up to the quartic order of the potential:
\be
\label{effdyn2}
\dot\alpha=\f \beta m
\,,\,\,
\dot\beta=\f{\hbar^{2}}{4m}\f1{\alpha^{3}}-\left(
V_{2}\alpha+\f{V_{4}}{2}\alpha^{3}+\f{V_{4}}{2}q^{2}\alpha+..
\right)
,
\ee
where we identify $\beta$ as the momentum conjugated to the position uncertainty $\alpha$. The equation for the evolution of $\beta$ is very similar to the equation for $\rd_{t}p$, up to the purely quantum extra-term ${\hbar^{2}}/{4m}{\alpha^{3}}$.

Another efficient method to derive the effective motion of a Gaussian wave-packet is to directly plug the Gaussian ansatz in the Schr\"odinger action, e.g. \cite{ABCV}. This ``Gaussian mini-superspace'', if one mimics the terminology used for reduced gravitational models in general relativity, gives the action driving the time evolution of the wave-packet\footnote{Here, we have put aside a total derivative contribution $\rd_{t}[-\hbar\gamma+\alpha\beta/2]$.}:
\be
\label{effaction}
S\big{[}\psi_{G}^{(q,p,A)}\big{]}
=
\int \rd t\,
\Big{[}
p\dot{q}+\beta\dot{\alpha}
-H_{\Delta}
-H_{V}
\Big{]}\,,
\ee
where we identify the two canonically conjugate pairs $(q,p)$ and $(\alpha,\beta)$ and the evolution is driven by the Hamiltonian $H=H_{\Delta}+H_{V}$.
The first contribution $H_{\Delta}$ comes from the Laplacian of the wave-function:
\be
\label{effham1}
H_{\Delta}
=
\f{\hbar^{2}}{2m}\int \rd x\, \pp_{x}\bpsi\pp_{x}\psi
=
\f{p^{2}}{2m}+\f{\beta^{2}}{2m}+\f{\hbar^{2}}{8m}\f1{\alpha^{2}}\,,
\ee
where we identify the free propagation terms in $p^{2}$ and $\beta^{2}$, as well as a conformal potential term in $\alpha^{-2}$ for the position uncertainty. The second contribution $H_{V}$ is the average of the potential evaluated on the Gaussian probability distribution:
\beq
\label{effham2}
H_{V}
&=&
\int \rd x\,V(x)|\psi|^{2}\\
&=&V(q)+\f{V_{2}}2\alpha^{2}+\f{V_{4}}8\alpha^{4}+\f{V_{4}}4q^{2}\alpha^{2}+..\nn
\eeq
We recognize the classical potential contribution $V(q)$ driving the evolution of the position expectation value, plus potential terms in $\alpha^{2n}$ driving the dynamics of the position uncertainty, plus potential terms coupling the position and momentum $(q,p)$ to the higher moments of the wave-function. The latter terms, such as the one in $q^{2}\alpha^{2}$, are due to anharmonic terms in the classical potential.
At the end of the day, the Hamilton equations arising from the effective action \eqref{effaction} are exactly the effective equations of motion \eqref{effdyn1} and \eqref{effdyn2} derived earlier directly from Schr\"odinger equation, showing the consistency of the whole approach.

\medskip

So, as we have explained up to now, focussing on the Gaussian wave-packet ansatz allows to study the dynamics of the (average) position and momentum and of the corresponding quadratic quantum uncertainty. One can extract from Schr\"odinger's equation, or even more directly from Schr\"odinger action principle, the effective equations of motion describing their dynamics.
Even though this approach does not see key quantum effects such as the discrete energy spectrum, it allows to recover the leading order classical motion plus quantum corrections. These corrections are  due to the non-trivial dynamics of the quantum uncertainty, i.e. the fluctuations of the shape of the wave-packet, and its non-trivial coupling to the classical position and momentum.

%
Here we would like to push this point of view further and analyze the higher than quadratic multipole moments of the wave-function. The goal is to describe the effective dynamics of those higher modes, which encode all the information about the shape of the wave-function.


\section{Tower of Higher Modes}

To study the dynamics of the wave-function's higher multipoles, we drop the hypothesis of a Gaussian wave-packet, study the evolution of an arbitrary wave-function, and focus on the case of a time-dependent harmonic potential,
\be
V(t,x)=\f12 m\om(t)^{2}x^{2}
\,,\quad
V_{2}=m\om^{2},\,\,V_{4}=0
\,,
\ee
where the coupling $\om(t)$ a priori depends on the time $t$.
At the classical level, the equations of motion become linear in $x$,
\be
\ddot{x}+\om(t)^{2}x=0
\,.
\ee
The time dependence of the coupling $\om(t)$ nevertheless leads to a non-trivial physical behavior, beyond the simpler harmonic oscillator.
At the quantum level, assuming a harmonic potential, thus truncating the Taylor expansion of the potential to second order, allows simplifies the dynamics of the system, by decoupling the evolution of the main position and momentum from the dynamics of the higher modes. Indeed, as one can see from the effective equations for the quantum motion \eqref{effdyn1},\eqref{effdyn2} or directly from the effective Hamiltonian \eqref{effham1},\eqref{effham2}, setting $V_{4}$ and higher potential terms to zero kills the effective couplings between $(q,p)$ and the quadratic uncertainties $(\alpha,\beta)$. This means that the position and momentum expectation values $(q,p)$ follows exactly the classical equations of motion. From this perspective, one might be tempted to say that the case of a harmonic potential is trivially quantized, since there is no feedback of quantum uncertainty on dynamics of main expectation values $\la \hx\ra$ and $\la \hp\ra$ corresponding to the classical degrees of freedom, or in short no quantum correction to the classical trajectories. Nevertheless, even if there is  no feedback of the higher order  fluctuations of the wave-functions on lower moments, those higher multipole moments characterizing the wave-function shape do exist, they are legitimate degrees of freedom in the theory and have their own non-trivial dynamics.

Let us thus study the evolution those higher wave-function modes and derive their effective dynamics. The  multipole moments of the wave-functions are the expectation values of the monomials $x^{a}p^{b}$. These observables include the position averages $\la x^{n}\ra$ of the probability distribution $|\psi(x)|^{2}$ and further allow to  probe the phase of the wave-function.
Using the Weyl symmetric ordering, they are defined by the corresponding polynomial operators:
\be
{\op}_{n,\ell}
=
(\widehat{p^{\ell}x^{n-\ell}})_{sym}
=
\f1{2^{\ell}}\sum_{k=0}^{\ell}\bino{k}{\ell} \hp^{k}\hx^{n-\ell}\hp^{\ell-k}
\,.
\ee
The expectation values $\la {\op}_{n,\ell}\ra$ define the multipole expansion of the wave-function, and fully characterize its shape and fluctuations. In particular, the three operators ${\op}_{2,\ell}$ for $n=2$ are the squared position $\hat{x}^{2}$, the squared momentum $\hat{p}^{2}$ and the dilatation generator $\hat{D}=\f12(\hx\hp+\hp\hx)$. They form a $\sl(2,\R)$ Lie algebra,
\be
\begin{array}{lcl}
[\hat{D},\hx^{2}]&=&-2i\hbar\hx^{2}
\,,\\
{[}\hat{D},\hp^{2}{]}&=&+2i\hbar\hp^{2}
\,,
\end{array}
\quad
[\hx^{2},\hat{p}^{2}]=+4i\hbar\hat{D}
\,,
\ee
and generate conformal transformations of the wave-functions. 
The $\sl(2,\R)$ Casimir operator simply reflects the Planck constant:
\be
\f12(\hat{x}^{2}\hat{p}^{2}+\hat{p}^{2}\hat{x}^{2})-\hat{D}^{2}
=
-\f{3\hbar^{2}}4
\,.
\ee
This $\sl(2,\R)$ Lie algebra is the key algebraic structure to integrate the case of a harmonic potential, even if time-dependent, since the Hamiltonian operator is then an $\sl(2,\R)$ element.
%
%
Their expectation values $\la {\op}_{2,\ell}\ra$  are the quadratic uncertainty observables in position and momentum, which characterize the Gaussian spread of a wave-packet.
The combination $\cC$ given by
\be
\cC\equiv \la \hat{x}^{2}\ra\la \hat{p}^{2}\ra- \la \hat{D}\ra^{2}
\ee 
is $\sl(2,\R)$-invariant, and thus is constant under evolution in a (time dependent) harmonic potential. Indeed, despite its non-trivial expression
\footnote{
A more usual observable is the uncertainty product,
\be
\Delta\equiv \delta_{x}\delta_{p}-\delta_{xp}
\quad\textrm{with}\,\,
\left|\begin{array}{lcl}
\delta_{x}&=&\la \hx^{2}\ra -\la\hx\ra^{2}\\
\delta_{p}&=&\la \hp^{2}\ra -\la\hp\ra^{2}\\
\delta_{xp}&=&\la \hx\hp\ra_{sym} -\la\hx\ra\la\hp\ra
\end{array}\right.
\,,
\ee
which is always bounded from below by the Robertson-Schr\"odinger inequality, $\Delta\ge \hbar^{2}/4$. This minimal value is reached for Gaussian wave-packets, $\Delta_{G}=\hbar^{2}/4$. Unlike $\cC_{G}$, it is directly a constant, it is a $\sl_{2}$-invariant, but it does not depend on the Gaussian parameters $(q,p,\alpha,\beta)$, and so does not provide a non-trivial constant of motion.
}
on Gaussian wave-packets,
\be
\label{ermakov}
\cC_{G}=(q\beta-p\alpha)^{2}+\f{\hbar^{2}q^{2}}{4\alpha^{2}}+\f{\hbar^{2}}{4}
\,,
\ee
the equations of motion \eqref{effdyn1}-\eqref{effdyn2} directly ensures that it is a constant of motion $\rd_{t}\cC_{G}=0$ for a time dependent harmonic potential. This is called the Ermakov-Lewis invariant.
It provides a time-independent measure of the quadratic uncertainty and can be shown to generate the symmetry under conformal time reparametrizations of classical and quantum mechanics in a harmonic potential \cite{Livine:2022vaj}.
Identifying similar invariants for more general potentials can be done in a systematic way, using canonical transformation methods,
\cite{Sarlet_1978,lewis1982direct,LEWIS1982133,Struckmeier_2001}, although their relation to symmetries does not seem yet to be understood.

Those three quadratic operators structure the whole space of higher multipole operators ${\op}_{n,\ell}$. 
Indeed, let us consider the action of the $\sl(2,\R)$ operators on all the monomial operators, by computing the commutators:
\beq
[\hD,\op_{n,\ell}]
&=&
+i\hbar\,(2\ell-n)\op_{n,\ell}
 \\
{[}\hp^{2},\op_{n,\ell}{]}
&=&
-2i\hbar\,(n-\ell)\op_{n,\ell+1}
\nn \\
{[}\hx^{2},\op_{n,\ell}{]}
&=&
+2i\hbar\,\ell\op_{n,\ell-1}
\nn
\eeq
Conformal operators do not change the overall degree $n$ of the monomial, but create shifts in $\ell$ at fixed $n$, which reflects the change of powers in $\hat{x}$ and $\hat{p}$. The dilatation operator simply reads the difference of powers  in $\hat{x}$ and $\hat{p}$, by attributing a conformal weight $-1$ to $x$ and $+1$ to $p$. The operator $\hat{p}^{2}$ increase the power in $p$ while lowering the power in $x$. The operator $\hat{x}^{2}$ does the reverse.
The consequence is that each degree $n$ defines a $\sl(2,\R)$ representation on the $(n+1)$-dimensional module $\cH_{n}=\bigoplus_{\ell=0}^{n}\C\op_{n,\ell}$. The dilatation generator $\hat{D}$ is naturally diagonalized in the $\ell$-basis. The operators $\hat{x}^{2}$ and $\hat{p}^{2}$ acting as lowering and raising operators. This organizes the moments of the wave-function in terms of their conformal dimensions and groups them into $\sl(2,\R)$ irreducible representations.

Since the Hamiltonian belongs to the $\sl(2,\R)$ Lie algebra, these commutators give the time evolution  for these higher order operators, $i\hbar \,\rd_{t}\op_{n,\ell}=[\op_{n,\ell},H]$, which turns into a system of differential recursion relations for their expectation values, 
\be
\label{eqntosolve}
\rd_{t}\la{\op}_{n,\ell}\ra
=
\f{n-\ell}{m}\la\op_{n,\ell+1}\ra-m\om^{2}\ell\la\op_{n,\ell-1}\ra
\,,
\ee
where the dynamics in a harmonic potential decouple the evolution of the wave-function moments according to their order $n$.

The main result of this paper is to provide the explicit solutions to these equations.
Let us come back to the classical equation of motion:
\be
\ddot{x}+\om(t)^{2}x=0\,,
\ee
and call $q_{1},q_{2}$ its two linearly independent solutions. Assuming, for the sake of normalization, that their Wronskian is set to 1, $W[q_{1},q_{2}]=q_{1}\dot{q}_{2}-q_{2}\dot{q}_{1}=1$, the second solution is given in terms of the first solution by:
\be
q_{2}(t)=q_{1}(t)\int^{t}\f1{q_{1}^{2}}
\,.
\ee
Then, we show that the space of solutions for $\la \op_{n,\ell}\ra $ is a $(n+1)$-dimensional vector space. More precisely, a set of linearly independent solutions for  $\la \op_{n,0}\ra$ at $\ell=0$ is given by the functions $q_{1}^{r}q_{2}^{n-r}$ for  integer exponents with $0\le r\le n$.
Solutions for $\ell>0$ are then extracted from the iterated time derivatives $\rd_{t}^{\ell}\la \op_{n,0}\ra$ according to \eqref{eqntosolve}.

The proof is a direct check that this produces explicit solutions to the equations of motion. Indeed, let us introduce the following ansatz labeled by the integer $r$, smaller or equal to $n$,
\be
\label{solutions}
\la \op_{n,\ell}\ra_{r}
=
\f{m^{\ell}}{\bino{r}{n}}
\sum_{a+b=\ell}
\f{\bino{r-a}{n-\ell}}{\bino{a}{\ell}}
\dot{q}_{1}^{a}\dot{q}_{2}^{b}q_{1}^{r-a}q_{2}^{n-r-b}
\,,
\ee
with $a\le r$ and $b\le n-r$, where the factorial factors can expanded for more clarity:
\be
\f{\bino{r-a}{n-\ell}}{\bino{r}{n}\bino{a}{\ell}}
=
\f{(n-\ell)!r!(n-r)!\ell!}{n!(r-a)!(n-r-b)!a!b!}
\,.
\ee
It is straightforward to check that, for each value of $r$, this expression satisfies the differential recursive equations of motion \eqref{eqntosolve}.

This illustrates that:
\begin{itemize}
\item For each multipole degree $n$, there is a $(n+1)$ space of solutions, i.e. $(n+1)$ independent integration constants encoding the evolution of the wave-functions. This means that, morally, the expectation values $\la \op_{n,\ell}\ra $ are all independent degrees of freedom.
This legitimatizes looking at that the moments $\la \op_{n,\ell}\ra $ as a complete set of observables encoding the amplitude and shape of the wave-function, instead of systematically using the Fourier modes of the wave-function.

\item In the present case of a harmonic oscillator, even if time-dependent, all the higher modes of the wave-function all beat together in some synchronous way. Indeed, they are all constructed as polynomials in the same two solution functions $q_{1}$ and $q_{2}$. This makes natural the proposal in \cite{Livine:2022vaj} to use quadratic (or higher) wave-function modes as clocks in which the classical system (i.e. the position and momentum expectation values) would beat regularly at a constant frequency.


\end{itemize}

At the mathematical level, the exact solutions \eqref{solutions} generalize the solutions for the quadratic position uncertainty. Indeed, $\alpha$ satisfies a non-linear second order differential equation,
\be
\ddot{\alpha}+\om(t)^{2}\alpha=\f{\hbar^{2}}{4m^{2}}\f1{\alpha^{3}}\,,
\ee
whose solutions are identified as the square-roots of second-order homogeneous polynomials in $q_{1}$ and $q_{2}$, explicitly:
\be
\alpha=\big{(}
\lambda q_{1}^{2}+\mu q_{2}^{2}+2\rho q_{1}q_{2}\big{)}^{\f12}
\,,
\ee
in terms arbitrary parameters $\lambda,\mu,\rho$ satisfying the normalization condition $\lambda\mu-\rho^{2}=1$.
Squaring this expression confirms the result \eqref{solutions} that the space of trajectories for $\la \cO_{2,0}\ra=\la \hx^{2}\ra=q^{2}+\alpha^{2}$ is spanned by the quadratic monomials $q_{1}^{2}$,  $q_{2}^{2}$ and $q_{1}q_{2}$.
The present analysis provides the general solutions to all higher multipoles of the wave-functions, beyond the quadratic moments.

Moreover, it leads to a natural generalization of the Ermakov-Lewis invariant \eqref{ermakov} for the quadratic uncertainty to constants of motion associated with higher multipoles. Indeed, using the $\sl_{2}$-invariant bilinear form on the $(n+1)$-dimensional irreducible $\sl(2,\R)$-representations, we consider the $\sl_{2}$-invariant combination of multipole moments at fixed integer $n$:
\be
\cC^{(n)}\equiv
\f12\sum_{\ell=0}^{n}(-1)^{\ell}\bino{\ell}{n}\la \cO_{n,\ell}\ra\la \cO_{n,n-\ell}\ra
\,,
\ee
which vanishes when $n$ is odd. These quantities all vanish classically, so they truly probe quantum effects.
One easily checks that these are all constants of motion of the time-dependent
\footnote{
Remember that the Hamiltonian is not conserved because of its explicit time dependence, $\rd_{t}H=\pp_{t}H\ne 0$ on classical trajectories. Similarly, quantities such as $\la \hp\ra^{2}+m^{2}\om^{2}\la \hx\ra^{2}$ are not constants of motion because the coupling $\om$ is explicitly time dependent, $\pp_{t}\om\ne 0$.
} quantum harmonic oscillator, $\rd_{t}\cC^{(n)}=0$. It matches the Ermakov-Lewis invariant for $n=2$, so that these are the higher order extension of this non-trivial invariant.

\medskip

This explicit and exact solution to the dynamics of the quantum time-dependent harmonic oscillator relies on the decoupling of the multipole moments of the wave-functions according to their degree $n$. Let us underline that this decoupling only works for a quadratic potential. Higher order potential will inevitably mix the $\sl(2,\R)$-modules with different $n$'s. This will lead to a more intricate quantum evolution with the high multipole moments  interfere with the lower modes. For instance, a quartic potential clearly couples the quadratic uncertainty to the expectation value of the position, as shown in equations \eqref{effdyn1} and \eqref{effdyn2}. In short, anharmonic potential terms couple the quantum dressing with the classical motion, and should reveal effects of higher order quantum fluctuations at the classical level.


\section{Discussion}

The primary purpose of this short paper was to look at the dynamics of the quantum uncertainty in standard non-relativistic quantum mechanics for Schr\"odinger equation with a possibly time-dependent potential.
This means focussing on the evolution of the observables $\la \hp^{\ell}\hx^{\ell'}\ra_{sym}$ instead of the Fourier modes of the wave-function. While Fourier modes are particularly relevant when looking at the scattering of pure momentum modes, the polynomial observables $\la \hp^{\ell}\hx^{\ell'}\ra_{sym}$ are the multipole moments of the wave-function. They are especially relevant from a  semi-classical viewpoint: they encode the quantum uncertainty and, more generally, describe the shape of the wave-function and its fluctuations.
This looks at quantum mechanics as classical mechanics dressed with quantum degrees of freedom. This perspective turns out to be fruitful to describe the Hamiltonian evolution of the wave-function, while putting aside the other key quantum processes of measurements and projections onto  eigenstates.

We first investigate the behavior of Gaussian wave-packets. This amounts to truncating the sequence of multipole moments by setting all moments with $\ell+\ell'>2$ to 0. We can then focus on the dynamics of the main position and momentum expectation values, $\la \hx\ra$ and $\la \hp\ra$, and on the quadratic moments, $\la \hx^{2}\ra$, $\la \hp^{2}\ra$ and $\la \hx\hp\ra_{sym}$. The position and momentum averages describe the classical trajectory, with possible semi-classical corrections, while the quadratic observables encode the leading order quantum uncertainty.
We extract the dynamics of those multipole moments directly from Schr\"odinger equation, or, using a technique from effective quantum mechanics, from an effective action obtained by evaluating Schr\"odinger action on the Gaussian wave-packet ansatz.
For a (time dependent) harmonic potential, the position and momentum follow the classical equations of motion, while the quadratic uncertainty has a decoupled non-linear dynamics. When one includes anharmonic potential terms, the classical equations of motion become non-linear and the consequence of this non-linearity at the quantum level is that one inevitably mixes the dynamics of the higher multipole moments with the lower ones, thereby non-trivially coupling the evolution of the classical degrees of freedom with the quantum uncertainty.
This reveals the quantum fluctuations of the wave-packets as legitimate physical degrees of freedom.

The second body of results deals with the extension of the analysis beyond the Gaussian wave-packet. While we consider a general wave-function without truncating its multipole expansion, we restrict ourselves to a harmonic potential, though possibly time dependent.
The equations of motion for the moments are derived directly from Schr\"odinger equation.
In this regime, the dynamics decouple the multipole moments layer by layer according to their order $n=\ell+\ell'$. Although each multipole moment is a legitimate independent degree of freedom, we show that the solutions to the equations of motion are all constructed as polynomials in the same two functions $q_{1}$ and $q_{2}$, which are the classical trajectories for the position $x$.

This provides a full exact solution to quantum mechanics in a time dependent quadratic potential (see e.g. \cite{Vachaspati:2018llo, Bojowald:2021cqg} for application to cosmology). This result further means that, despite the time dependence of the potential, all the moments tick in the same way.
Let us nevertheless underline that the higher moments evolve as higher powers of the classical motions, so involve higher harmonics of the frequencies driving the classical system.
In spite of this subtlety, this analysis makes natural the result of \cite{Livine:2022vaj} that the evolution of the quantum uncertainty can be used to define an intrinsic clock to describe the evolution of the classical position and momentum, a kind of intrinsic rest frame where a time-dependent harmonic well loses its time dependence and defines a constant frequency for the system.
Then quartic and higher polynomial terms in potential will mix the layers of wave-function multipoles together and create a complex dynamics, where quantum degrees of freedom can reveal themselves through deviations of the classical trajectory.

This work should be considered as a first step in studying further the dynamics of the quantum uncertainty. It would, of course, be very interesting to analyze the evolution of the quadratic uncertainty (i.e. the squeezing) of wave-packets, and of more general wave forms, for physically relevant potentials, which includes anharmonic terms.
{
Focussing for instance on a (possibly time-dependent) quartic potential. One would look at the dynamics of the the first and second order moments generated by the effective Hamiltonian \eqref{effham2}, which entangles the classical motion with the quantum uncertainty, and check it against numerical simulations, in order to test the limits of validity of this effective approach.
Then moving on to the dynamics of the higher moments, the quartic potential will couple moments of different order, $[\hx^{4},\cO_{n,\ell}]\propto \cO_{n+2,\ell-1}$, and we will need to study the validity of truncations where one would be neglect modes of order $n>n_{max}$. This is very similar in logic as a renormalization scheme, where we would analyze how the predicted evolution depends on the truncation level $n_{max}$.
Finally, the $sl_{2}$ invariants, $\cC^{(n)}$, which reproduce in particular the Ermakow-Lewis invariant for $n=2$, and which are constants of motion for a time-dependent quadratic potential, can be used to track the precision of numerical simulations in that case. But a non-trivial evolution, in the case of a higher order potential, would signal a significant deviation from the harmonic behavior and large quantum fluctuations of the wave-packet shape due to higher moments. Understanding how it works precisely would require a numerical study.
}

Beyond more general potentials, one could also look at the effective dynamics of (the squeezing of) wave-packet superpositions (see e.g. \cite{Jalabert_2001} for work on decoherence), for example, to apply to experimental tests of the Di\'osi-Penrose gravitationally-induced 
\footnote{Let us point out, by a dimensional argument, that the fundamental unit for the relative uncertainty in momentum and position, $[{\delta x}/{\delta p}]=[G][c^{-3}]$,
is Newton's universal constant of gravitation, so one naturally expects gravitation to deeply affect the dynamics of the squeezing of quantum states, at least in the regime of very small uncertainty ratio.
{This remark is actually the starting point for the proposal of generalized uncertainty principles (GUPs) in quantum gravity phenomenology.}
}
decoherence \cite{Penrose:2000ic} and of the gravity-quantum interplay \cite{Belenchia:2019gcc,Westphal:2020okx,Christodoulou:2022mkf,Biswas:2022qto}, and more particularly to the evolution of the wave-packet shape and uncertainty during  scattering events. In order to have an interaction between wave-packets and scattering, one needs to add a self-interaction term to the Schr\"odinger equation. For instance, one would look at the interaction and scattering for a coherent superposition of two wave-packets for a Non-Linear Schr\"odinger equation
of their quantum uncertainties.
{
To be more explicit, let us consider 1d NLSE with a $|\psi|^{4}$ self-interaction term,
\be
i\hbar \pp_{t}\psi=-\gamma\pp_{x}^{2}\psi+2\ka |\psi|^{2}\psi\,,
\quad\textrm{with}\,\,\gamma={\hbar^{2}}/{2m}\,.
\ee
Although the momentum expectation value $\la p\ra$ is still conserved due to translational invariance, this is not the case anymore for the momentum spread $\la p^{2}\ra$ due to the non-linearity, so that we can not extend the present analysis in a straightforward manner. It is however known that this is an integrable model \cite{Zakharov:1974zf}, implying that the evolution could be exactly solved in principle. In particular, there exists an infinite tower of conserved charges $Q_{n}=\int \rd x\,\bpsi u_{n}$, given by a recursion relation \cite{Segur:1976:ASCa,Faddeev1987HamiltonianMI, Pritula_2002,LocalConsLawsNLS},
\be
u_{1}=\psi\,,\quad
u_{n}=-i\pp_{x}u_{n-1}+\gamma^{-1}\ka\sum_{k=1}^{n-2}u_{k}u_{n-1-k}
\,.
\ee
The first charge is the number of particles $Q_{1}=\int |\psi|^{2}$, the second charge is the momentum, the third charge is the NLS Hamiltonian, and then we get all non-linear conserved extension of the powers of the momentum $\la p^{n}\ra$.
Not only are those charges relevant to esnure the stability of numerical simluations of the NLSE, but the hope is also to use them to integrate the motion for powers of the position and more generally all moments $\la x^{a}p^{b}\ra$. This will reveal the dynamics of quantum uncertainty for a self-interacting wave-function,
}
all  in the purpose of understanding better the quantum-to-classical transition from a field theory perspective.

%
%

%





\begin{thebibliography}{0}

\bibitem{Bojowald:2005cw}
\Name{M.~Bojowald \and A.~Skirzewski}
\REVIEW{Rev. Math. Phys.}{18}{2006}{247}

\bibitem{Bayta2019}
\Name{B.~Bayta{\c{s}} \and M.~Bojowald \and and S.~Crowe}
\REVIEW{Physical Review A}{99}{2019}{}

\bibitem{Bojowald:2022lbe}
\Name{M.~Bojowald}
\REVIEW{J. Phys. A}{55}{2022}{50}

\bibitem{Heller}
\Name{E.J.~Heller}
\REVIEW{The Journal of Chemical Physics}{64}{2008}{63-73}

\bibitem{ABCV}
\Name{F.~Arickx \and J.~Broeckhove \and W.~Coene \and P.~Van~Leuven}
\REVIEW{International Journal of Quantum Chemistry}{30}{1986}{471-481}

\bibitem{Pattanayak_1994}
\Name{A.~K. Pattanayak \and W.~C. Schieve}
\REVIEW{Physical Review E}{50}{1994}{5,3601-3615}


\bibitem{Prezhdo}
\Name{O.~V. Prezhdo \and Y.~V. Pereverzev}
\REVIEW{Journal of Chemical Physics}{113}{2000}{6557-6565}

\bibitem{Blum:1995wi}
\Name{T.~C. Blum \and H.~T. Elze}
\REVIEW{Phys. Rev. A}{53}{1996}{3123}

\bibitem{Livine:2022vaj}
\Name{E.~R. Livine}
\Book{http://arXiv.org/abs/2212.09442}

\bibitem{Kanasugi}
\Name{H.~Kanasugi}
\REVIEW{Progress of Theoretical Physics}{97}{1997}{617--633}

\bibitem{PhysRevResearch.2.043162}
\Name{J.~G. Muga \and S.~Mart\'{\i}nez-Garaot \and M.~Pons \and M.~Palmero \and A.~Tobalina}
\REVIEW{Phys. Rev. Res.}{2}{2020}{043162}

\bibitem{fiore2022timedependent}
\Name{G.~Fiore}
\Book{http://arXiv.org/abs/2205.01781}

\bibitem{PhysRevA.43.404}
\Name{C.~F. Lo}
\REVIEW{Phys. Rev. A}{43}{1991}{404--409}

\bibitem{Belenchia:2019gcc}
\Name{A.~Belenchia \and R.~M. Wald \and F.~Giacomini \and E.~Castro-Ruiz \and V.~Brukner \and
  M.~Aspelmeyer}
\REVIEW{Int. J. Mod. Phys. D}{28}{2019}{1943001}

\bibitem{Westphal:2020okx}
\Name{T.~Westphal \and H.~Hepach \and J.~Pfaff \and M.~Aspelmeyer}
\REVIEW{Nature}{591}{2021}{7849}

\bibitem{Christodoulou:2022mkf}
\Name{M.~Christodoulou \and A.Di~Biagio \and M.~Aspelmeyer \and V.~Brukner \and C.~Rovelli \and  R.~Howl}
\REVIEW{Phys. Rev. Lett.}{130}{2023}{100202}

\bibitem{Biswas:2022qto}
\Name{D.~Biswas \and S.~Bose \and A.~Mazumdar \and M.~Toro\v{s}}
\Book{http://arXiv.org/abs/2209.09273}

\bibitem{Sarlet_1978}
\Name{W.~Sarlet}
\REVIEW{Journal of Physics A}{11}{1978}{843-854}

\bibitem{lewis1982direct}
\Name{H.R. Lewis \and P.~Leach}
\REVIEW{Journal of  Mathematical Physics}{23}{1982}{2371-2374}

\bibitem{LEWIS1982133}
\Name{H.R. Lewis \and P.~Leach}
  \Editor{A.~Bishop, D.~Campbell, and B.~Nicolaenko}
 \Book{Nonlinear Problems: Present and Future}
 \Vol{61}
  \Publ{  North-Holland Mathematics Studies}
  \Year{1982}
  \Page{133-145}.
%

\bibitem{Struckmeier_2001}
\Name{J.~Struckmeier \and C.~Riedel}
\REVIEW{Physical Review E}{64}{2001}{}

\bibitem{Vachaspati:2018llo}
\Name{T.~Vachaspati \and G.~Zahariade}
\REVIEW{Phys. Rev. D}{98}{2018}{065002}

\bibitem{Bojowald:2021cqg}
\Name{M.~Bojowald \and B.~Jones}
\REVIEW{JCAP}{11}{2021}{037}

\bibitem{Jalabert_2001}
\Name{R.A. Jalabert \and H.M. Pastawski}
\REVIEW{Physical Review Letters}{86}{2001}{2490--2493}

\bibitem{Penrose:2000ic}
\Name{R.~Penrose}
 \Book{9th Marcel Grossmann Meeting on Recent
  Developments in Theoretical and Experimental General Relativity, Gravitation
  and Relativistic Field Theories (MG 9)}
 \Vol{7}
  \Publ{  North-Holland Mathematics Studies}
  \Year{2000}
  \Page{3--6}.

\bibitem{Zakharov:1974zf}
\Name{V.~E. Zakharov \and S.~V. Manakov}
\REVIEW{Teor. Mat. Fiz.}{19}{1974}{332--343}

\bibitem{Segur:1976:ASCa}
\Name{H.~Segur \and M.~J. Ablowitz}
\REVIEW{Journal of  Mathematical Physics}{17}{1976}{  710--713}

\bibitem{Faddeev1987HamiltonianMI}
\Name{L.~D. Faddeev \and L.~A. Takhtajan}
  \Book{Hamiltonian methods in the theory of  solitons}
  \Publ{Springer}
  \Year{1987}

\bibitem{Pritula_2002}
\Name{G.~M. Pritula \and V.~E. Vekslerchik}
\REVIEW{Inverse Problems}{18}{2002}{1355--1360}

\bibitem{LocalConsLawsNLS}
\Name{J.~Barrett}
 \Book{The local conservation laws of the nonlinear Schr\"odinger  equation}
  \Publ{PhD thesis (Oregon State University)}
  \Year{2013}

\end{thebibliography}

\end{document}